\documentclass[bibyear]{aa}  
\usepackage{graphicx}
\usepackage{txfonts}
\usepackage[colorlinks=true, 
urlcolor=blue, 
citecolor=blue, 
linkcolor=blue]{hyperref}
\usepackage{natbib}
\usepackage{booktabs}
\usepackage{xcolor}
\usepackage[switch]{lineno}
\usepackage{orcidlink}

\begin{document} 

   \title{Core-collapse supernova parameter estimation with the upcoming Vera C. Rubin Observatory}
   \titlerunning{CCSN parameters with LSST}
   \subtitle{}

    \author{
    Andrea Simongini$^{\orcidlink{0009-0000-3416-9865}}$,$^{1,2}$\thanks{E-mail: andrea.simongini@inaf.it}
    Fabio Ragosta$^{\orcidlink{0000-0003-2132-3610}}$,$^{3,4}$ 
    Irene Di Palma$^{\orcidlink{0000-0003-1544-8943}}$ $^{5,6,1}$ \and
    Silvia Piranomonte$^{\orcidlink{0000-0002-8875-5453}}$$^{1}$ 
    }
    \institute{
    $^{1}$ INAF - Osservatorio Astronomico di Roma, Via di Frascati 33, I-00078 Monteporzio Catone, Italy\\
    $^{2}$ Università Tor Vergata, Dipartimento di Fisica, Via della Ricerca Scientifica 1, I-00133 Rome, Italy \\
    $^{3}$ Dipartimento di Fisica “Ettore Pancini”, Università di Napoli Federico II, Via Cinthia 9, 80126 Naples, Italy \\ 
    $^{4}$ INAF - Osservatorio Astronomico di Capodimonte, Via Moiariello 16, I-80131 Naples, Italy\\
    $^{5}$ Universitá La Sapienza, Dipartimento di Fisica, Piazzale Aldo Moro 2, I-00185 Rome, Italy\\
    $^{6}$ INFN, Sezione di Roma, 00133 Rome, Italy 
    }
    \authorrunning{A. Simongini et al.}

   \date{Received 25 March, 2025; accepted 29 May, 2025}

   \abstract{The Vera Rubin Observatory’s Legacy Survey of Space and Time (LSST) is expected to revolutionize time-domain optical astronomy as we know it. With its unprecedented depth, capable of detecting faint sources down to r $\sim$ 27.5 mag, the LSST will survey the southern hemisphere sky, generating nearly 32 trillion observations over its nominal 10-year operation. Among these, approximately 10 million will be supernovae (SNe), spanning a wide range of redshifts, with an expected rate of 6.8 10$^{-5}$ SNe Mpc$^{-3}$  yr$^{-1}$ . These observations will uniquely characterize the SN population, enabling studies of known and rare SN types, detailed parameterization of their light curves, deep searches for new SN progenitor populations, the discovery of strongly lensed SNe, and the compilation of a large, well-characterized sample of superluminous SNe. We analyzed a sample of 22663 simulations of LSST light curves for core collapse supernovae (CCSNe). The explosions were modeled using the radiative transfer code \texttt{STELLA}, and each event was provided with a value of redshift, extinction, cadence, explosion energy, nickel yield, and progenitor mass. We analyzed this dataset with the software \texttt{CASTOR}, which enables the reconstruction of synthetic light curves and spectra via a machine learning technique that allows one to retrieve the complete parameter map of a SN. For each parameter we compared the observed and the true values, determining how LSST light curves alone will contribute to characterize the progenitor and the explosion. Our results indicate that LSST alone will not suffice for a comprehensive and precise characterization of progenitor properties and explosion parameters. The limited spectral coverage of LSST light curves (in most cases) does not allow for the accurate estimation of bolometric luminosity, and consequently, of the explosion energy and nickel yield. Additionally, the redshift-absorption degeneracy is difficult to resolve without supplementary information. These findings suggest that for the most interesting SNe, complementary follow-up observations using spectrographs and optical facilities (particularly in the infrared bands) will be essential for accurate parameter determination.

} 

   
   \maketitle


\section{Introduction}

    The upcoming era of the Vera C. Rubin Observatory's Legacy Survey of Space and Time (VRO-LSST) will revolutionize the field of transient astrophysics as we know it today \citep{Ivezic2019}. Located on Cerro Pachón in northern Chile, the VRO will be a large, wide-field, ground-based telescope specifically designed to take highly detailed wide-field snapshots of the entire southern hemisphere sky, repeated every few nights, with a not-so-common ability to see very faint sources (r $\sim$ 27.5 mag). Major science programs will be accomplished during the LSST survey, a deep-wide-fast survey that represents the latest culmination of the technological advancement that has characterized the field of large surveys over the past 20 years. The LSST survey will occupy about 90 percent of the total observing time, observing a 18.000 deg$^2$ region about 800 times (summed over all its six bands, i.e., ugrizy) during the 10 years of operations. The final database will include about 32 trillion observations of 20 billion galaxies and a comparable number of individual stars \citep{Ivezic2019}. The strength of the VRO lies in its compact size and the rigidity of its telescope mount, as well as its constantly moving crawling dome that anticipates the telescope's next position. These features will enable the telescope to change position rapidly, with an unprecedented velocity for a facility of its size, approximately 5 seconds \citep{LSST2009, Ivezic2019}.

    The LSST survey is expected to increase the number of detected transient events by more than two orders of magnitude compared to current surveys, identifying approximately 10 million changes in the sky each night \citep{graham2019lsst, Ivezic2019}. Thanks to its unique capability to simultaneously provide large-area coverage, dense temporal coverage, accurate color information, good image quality, and rapid data reduction and classification, LSST will represent a milestone in the transient astronomy field. Since LSST extends the time–volume–color space 50–100 times over current surveys, it will facilitate new population and statistical studies and also the discovery of new classes of objects. Among the millions of detected transients, it is anticipated that around 10 million supernovae (SNe) will be observed over the course of a decade, including approximately 14.000 objects with exceptionally detailed light curves, each with more than 100 photometric points across five bands \citep[see][for a detailed description]{Hambleton2023}.

    This vast dataset will provide an unparalleled characterization of the SN population and, in particular, of core-collapse SNe \citep[CCSNe; see e.g.][]{Filippenko1997, Branch2017, Hambleton2023}. It will enable studies of known and unusual SN populations and parameterization of their light curves, a deep search for new populations of SN progenitors, the discovery of strongly lensed SNe, and the identification of a large, well-characterized sample of superluminous SNe \citep[SLSNe; for a review see][]{Moriya2018}. 
    
    During the nominal 10-year survey, CCSNe will be discovered with a rate of $6.8 \times 10^{-5} \mbox{SNe} \, \mbox{Mpc}^{-3} \, \mbox{yr}^{-1}$. Among these, 70 percent are expected to be of type II-P, 15 percent of type Ib/c, 10 percent of type II-L, and only 0.05 percent of type IIn \citep{LSST2009}. Additionally, almost 200 thousand SLSNe will be discovered, boosting the current statistics in an unprecedented way. In particular, according to \citet{Rau2009}, the universal rate of SLSNe is 10$^{-7}$ Mpc$^{-3}$ yr$^{-1}$, with only $\sim$ 7 events discovered under 200 Mpc in the last 15 years \citep[e.g.][]{Bose2018}. 
    
    This work aims to understand the scientific impact of the LSST in the field of CCSNe, by evaluating the efficiency of parameter estimation from LSST simulated optical data. The correct parametrization of a SN can help in several ways: modeling the explosion mechanism and the powering sources at play, understanding stellar populations and consequently constraining the stellar evolution, and inferring the nucleosynthesis and feedback outcomes of the explosion. We analyze LSST simulated light curves with the open-access software \texttt{CASTOR} \citep{Simongini2024}, constraining the distance, the extinction, the explosion energy, the mass of nickel, and the progenitor's star mass. In view of upcoming observations, we evaluate with a statistical approach how well LSST data alone will allow us to characterize the parametric map of SNe.
    
\section{Data sample}

    We have used the simulations of red supergiant explosions by \citet{Moriya2023}. They presented a grid of 228016 synthetic explosions of typical type II SNe, based on progenitor models by \citet{sukhbold2016core}. The pre-SN evolution was modeled using the \texttt{KEPLER} code \citep{weaver1978presupernova}, while \citet{Moriya2023} utilized the open-source, one-dimensional, multifrequency radiative transfer code \texttt{STELLA} \citep{Blinnikov1998, Blinnikov2006, Blinnikov2004, Baklanov2005} to simulate the resulting light curves and photospheric evolution.

    In total, we collected a subset of the initial grid that contains the synthetic explosions of 22663 different SNe. Each simulation was provided with the information on the progenitor star's mass (M$_{pro}$), the explosion energy (E), the mass of nickel produced by radioactive decay post-explosion (M$_{ni}$), the mass-loss rate before the onset of the explosion (M$_{loss}$), the radius of the circumstellar material (R$_{csm}$), and the wind structure parameter ($\beta$). In particular, the parameter space covered by our subset is five progenitor masses (10, 12, 14, 16, and 18 $M_\odot$), ten explosion energies ($0.5, 1.0, 1.5, 2.0, 2.5, 3.0, 3.5, 4.0, 4.5, \text{and}\, 5.0 \times 10^{51}$ erg), and two masses of nickel (0.001 and 0.01 $M_\odot$). 

    Here, we organize the data into four levels of characterization. The first data level, $L_0$, contains the simulations. The second data level, L$_1$, was obtained by simulating a realistic distribution of SNe across the three-dimensional sky, adding redshift, extinction, and cadence information to each simulated explosion. This level models the anticipated SN distribution over the projected 10-year LSST survey \citep{LSST2009}.

\begin{figure}
    \includegraphics[width=1\columnwidth]{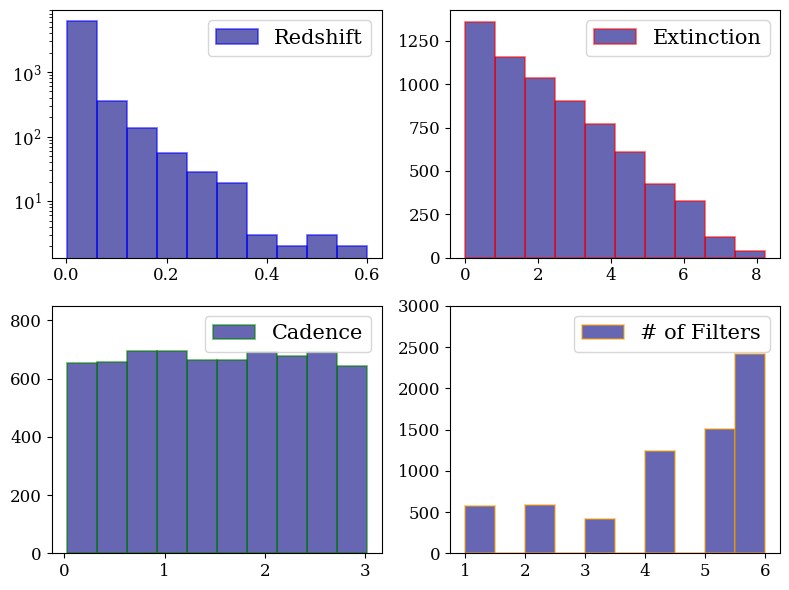}
    \caption{Distribution of redshift (top left panel), extinction (top right), cadence (bottom left), and number of filters (bottom right) of the 6730 SNe in the L$_2$ level of data.}
    \label{fig:database}
    \end{figure}

    The third level, L$_2$, filters out data points that fall outside the magnitude range defined by the nominal LSST's magnitude limit and saturation limit for each band. Specifically, it retains only points within the ranges 14.7 $-$ 23.8 in u, 15.7 $-$ 24.5 in g, 15.8 $-$ 24.03 in r, 15.8 $-$ 23.41 in i, 15.3 $-$ 22.74 in z, and 13.9 $-$ 22.96 in y \citep{LSST2009}. Additionally, only objects with at least ten detectable points within these limits in their light curves are retained. Following this filtering process, 6730 SNe remain ($\sim$ 30 percent of all simulations). The distributions of redshift, absorption, cadence, and peak magnitude for these L$_2$ SNe are shown in Fig~\ref{fig:database}. Note that nearly 19 percent\footnote{Note that, from now on, all percentages are given with respect to the remaining sample in the $L_2$ level, i.e., 6730 objects.} of the remaining sample is flagged as “saturated,” indicating that at least one point exceeded the saturation limit and was subsequently filtered out. In these cases, the final light curves will exhibit a gap near maximum brightness in at least one filter.
            
    The final data preparation step involves Gaussian process (GP) interpolation of the light curves, creating the L$_3$ dataset, which serves as the foundation for parameter estimation. The GP is a strong data-driven and nonparametric tool for interpolating the great morphology of SNe light curves \citep{Rasmussen2004}. In particular, due to its model-independent nature, it allows for an unsupervised approach, which is particularly effective for huge samples of data.   
    We performed GP interpolation with \texttt{CASTOR}, sampling data between the first and the last available data point in every filter to avoid nonphysical conditions. The kernel used is in the form 
    \begin{equation}
    k(x) = A\left(1+ \frac{\sqrt{3}x}{\sigma}\right)\mbox{exp}\left(-\frac{\sqrt{3}x}{\sigma}\right),
    \label{eq:kernel}
    \end{equation}
    where $A$ is an amplitude factor and $\sigma$ is the lengthscale at which the correlations between two measurements, $x$ and $x'$, are significant. To account for the differences in cadence, we set a mixture of kernels, obtained changing the lengthscale to the mean, maximum, and minimum cadence of each light curve, while the amplitude was set as the mean value of magnitude. If the SN being studied had a data gap exceeding ten times the sampling rate due to surpassing the saturation limit, we applied the GP in segments, leaving that gap unfilled. 
    
    \section{Simulation analysis with \texttt{CASTOR}}

    The simulation analysis begins at data level L$_3$, with the GP interpolated light curves. We used \texttt{CASTOR} again for the following steps. \texttt{CASTOR} is well suited for doing an analysis based solely on light curves, as it leverages the construction of synthetic spectra to derive the parameters of an event when no spectral data are available.
    
    \subsection{Synthetic spectra}

\begin{figure}
    \includegraphics[width=1\columnwidth]{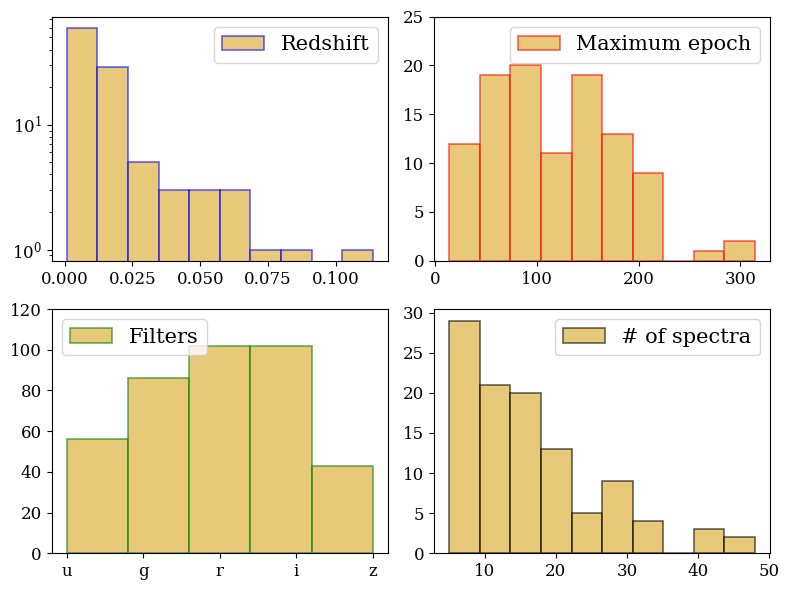}
    \caption{Distribution of redshift (top left), maximum available spectral epoch (top right), available filters (bottom left), and number of available spectra (bottom right) of the training set. }
    \label{fig:training}
    \end{figure}

     The first step of the analysis was the light curve comparison, which was performed by means of a chi-square test between the GP interpolated light curves and a catalog of 124 CCSNe \citep[][and references therein]{Simongini2024, castor_zenodo_2024}. Each of these SNe has been spectrophotometrically observed across a range, from near-UV to infrared, with more than five optical spectra available, and they are located at different distances. Among the various available filters, the ugriz-SLOAN filters are the most comparable to the LSST ones in terms of throughput \citep{Gunn1998}. Therefore, we considered only the SNe that had observations in these filters, reducing the catalog of reference SNe to a total of 106. Fig.~\ref{fig:training} shows the distribution of redshift, number of available spectra, maximum available epoch, and number of light curves for the 106 SNe of the training set. Comparing the database (Fig.~\ref{fig:database}) with the training set, it is immediately clear that the upper-end distribution of redshift does not have a counterpart in the training set, from z = 0.1 up to z = 0.6. This is mainly due to the advancement in technical capabilities that only allowed one to detect really far SNe in recent years and to the fact that, generally, farther SNe receive less attention and optical follow-ups, effectively producing a lack of publicly available data. Another important difference is the availability of filters: less than half of the SNe from the training set have enough data in the $u$ and $z$ filters, and none in the $y$ filter. However, the lack of light curves does not affect the synthetic spectra reconstruction, but it can have a relatively significant weight in the comparison process. Indeed, the aim of this process is to find the SN out of the training set (the so-called “reference SN”) whose light curves best resemble those of the SN of study. The observed spectral data of the reference SN, combined with the L$_3$ light curves converted into flux densities, are then used to build synthetic spectra. As is described in \citet{Simongini2024}, synthetic spectra are built between the time of explosion and the maximum available epoch of the reference SN via two-dimensional GP interpolation. The wavelength coverage of the final synthetic spectra is also dependent on the mean coverage of the reference SN. We used a combination of two Matern kernels with $\nu=1.5$, with the time lengthscales fixed as the minimum and maximum sampling step and the wavelength lengthscale fixed at 70\, $\AA$. In total, we built 20 spectra for every SN, to cover the region around the peak without having an effect on the computation time.

    \subsection{Parameters}

    A detailed description of how each parameter was estimated can be found in \citet{Simongini2024}, whereas details of the used version of the software are in \citet{castor_zenodo_2024}. All parameters were estimated from the L$_3$ light curves and the synthetic spectra and relied on specific physical assumptions; in particular, spherical symmetry of the explosion, perfect adiabaticity at the peak of luminosity, complete conversion of explosion energy into kinetic energy with the canonical partition between neutrinos (99.9 percent of the total energy) and photons (0.1 percent of the total energy), and perfect conservation of mass and canonical nucleosynthesis processes. Moreover, the distance of each object was estimated via a Hubble law, with $H_0 = 70$ km s$^{-1}$ Mpc$^{-1}$. We emphasize that this may not be the best approximation of distance of local galaxies, due to peculiar velocity effects with the consequence of higher errors of reconstruction at short distances. However, since the true distance values were also obtained under the same assumption, this error only propagates to the other parameters without affecting the distance comparison itself. Another important assumption is that two photometrically similar SNe have spectra that behave in the same way, thereby exhibiting the same or at least overlapping spectral features.
  
    Among the estimated parameters, we can distinguish between independent and degenerate parameters. The first class depends solely on data, without suffering from error propagation from other estimated parameters, and thus they are the most reliable for a direct comparison with the true value. Among these, redshift and velocity of the ejecta are directly estimated from the relative shift and the half-width of spectral lines, respectively, while extinction and time of maximum luminosity are taken directly from the synthetic light curves. Specifically, the extinction is estimated as the difference between a blue and a visible filter at maximum luminosity and is subsequently converted into absorption using the filter-specific corrections provided by \citet{McCall2004}. However, when the event is observed with only one or two widely separated filters (e.g., a blue and a red filter), a direct estimation from light curves is not possible. In such cases, we use the model from \cite{Cardelli1989} with $R_V=3.1$ averaging the extinction over the available filters.  
        
    The degenerate parameters, on the other hand, are affected by error propagation, depending on both independent parameters and other degenerate parameters. The bolometric luminosity was estimated from the absolute magnitude light curves, and therefore depends on both distance and extinction. However, when observations were limited to only one filter, we estimated the bolometric luminosity by integrating synthetic spectra. This approach allows for broader filter coverage but introduces a different temporal coverage, as it depends on the latest available epoch for the reference SN. Note that this is not the preferred approach because we aim to characterize any event using the maximum information coming from LSST light curves and not from the synthetic spectra. The energy was obtained by integrating the luminosity between the explosion epoch and the time of maximum luminosity, while the mass of nickel was estimated by applying the model from \citet{Lusk2017} on the linear decay of the bolometric luminosity. Consequently, it is not possible to estimate the mass of nickel when the light curves lack of observations in the linear decay stage. This affects almost $\sim$ 20 percent of all simulations in the L$_3$ level of analysis, for which data points in the decay phase are either not simulated or under the sensitivity limit of LSST. In these cases, the mass of nickel is simply set to zero and not taken into account for the general considerations of efficiency. Finally, the mass of the ejecta depends on both the energy and velocity of the ejecta via a virial theorem \citep{Arnett1982} and gives direct access to the mass of the progenitor. In particular, \texttt{CASTOR} gives an interval of equally probable masses of the progenitor star, accounting for all the possible values of the mass of the remnant. In this work, we present our results using the two bounds of this interval, which account for the minimum mass of a neutron star (NS) and the maximum mass of a black hole (BH) remnant.

\section{Figures of merit}

    We reconstructed the parametric map for each simulation and highlight two figures of merit that, from a statistical perspective, aim to quantify the main challenges in parameter estimation using LSST data for a large sample. Our analysis focuses on five key parameters: distance, extinction, energy, nickel mass, and progenitor mass. 

    \subsection{Kullback–Leibler divergence} 

\begin{figure*}
    \includegraphics[width=1\textwidth]{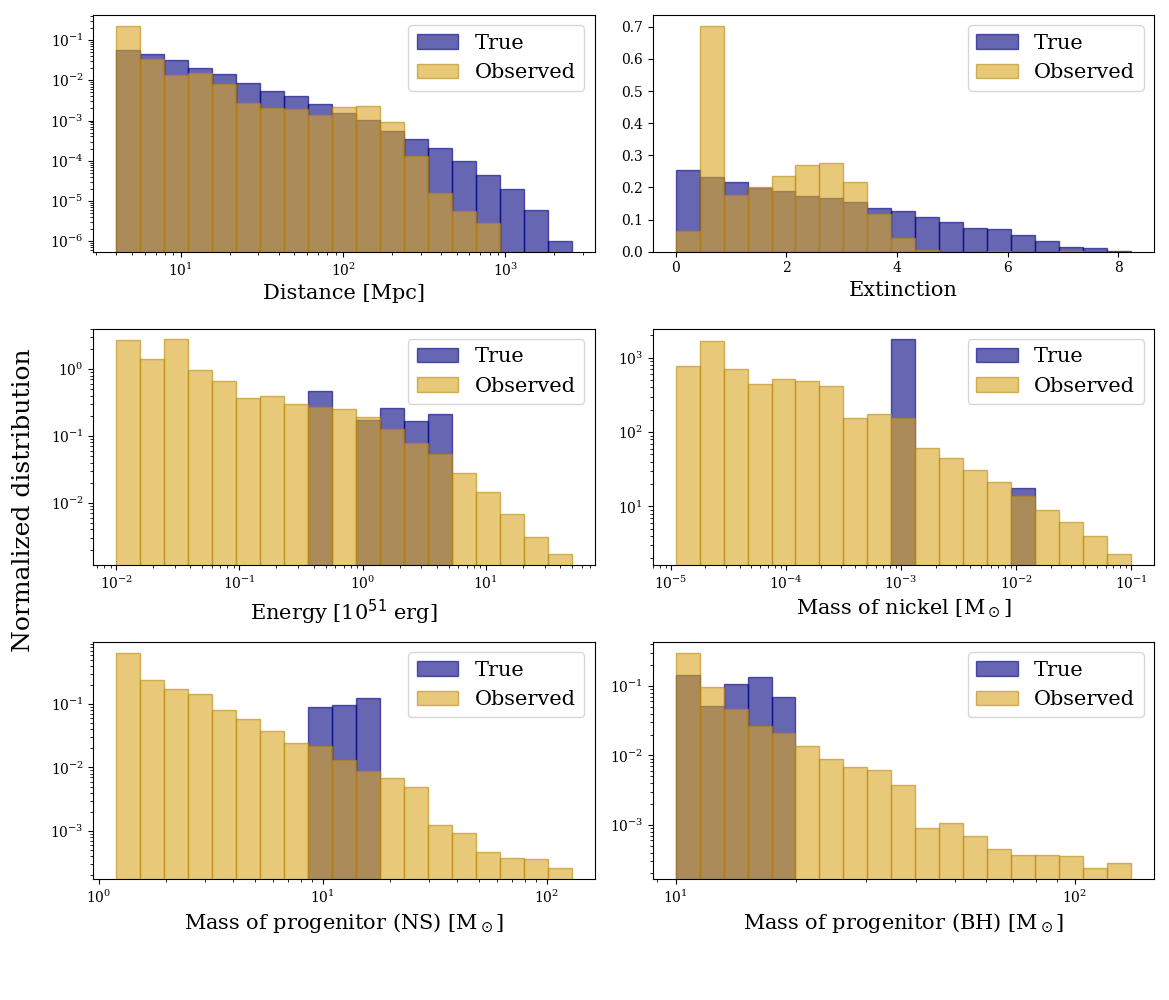}
    \caption{Comparison between the true (dark blue) and the observed (gold) distribution for each estimated parameter. Except for the extinction one, every plot is log scaled and the distributions are binned logarithmically. The normalization (y axes) changes for each distribution. Note that, because the bins are logarithmically distributed, the density is estimated differently for each bin.
    }
    \label{fig:comparison}
    \end{figure*}

    We compared the distributions of true and estimated parameters using the Kullback–Leibler \citep[KL;][]{Kullback1951} divergence: 
    \begin{equation}
        D_{KL}  = \sum_i P_i \mbox{log}_2 \left(\frac{P_i}{Q_i}\right). 
    \end{equation}
    This metric provides a nonsymmetric measure of the difference between two distributions: the true distribution, $P$, and its approximation, $Q$. The KL divergence approaches 0 as the two distributions become more similar and exceeds 1 when the differences are significant. In this work, $P$ and $Q$ represent the true (known from the initial conditions of the simulations) and estimated distributions of each parameter, respectively. By using KL divergence, we treat the dataset as a whole rather than focusing on individual discrepancies between SNe. Results are presented in Table~\ref{tab:results} and Fig.~\ref{fig:comparison}. Note that in order to maintain a statistical consistency we removed the outliers, defined on the basis of physical common values and statistical weight ($E > 50 \times 10^{51}$ erg, $E < 0.01 \times 10^{51}$ erg, $M_{pro} > 130 M_\odot$, $M_{ni} > 0.1 M_\odot$ and $M_{ni} < 10^{-5} M_\odot$ ). No limits were imposed on distance or extinction. 

    \begin{table}[h!]
    \centering
    \caption{Results of the comparison between the true and the estimated distributions.}
    \label{tab:peak}
    \begin{tabular}{l|cc} 
    \toprule 
    Parameter               & D$_{KL}$ & FoM \\  
    \midrule 
    Distance                & 1.20        & 2.75    \\
    Extinction              & 0.53        & 1.69    \\
    Energy                  & 1.37        & 2.61    \\
    Mass of nickel          & 2.15        & 15.9    \\ 
    Mass of progenitor (NS) & 0.74        & 0.84    \\ 
    Mass of progenitor (BH) & 0.18        & 0.52    \\ 
    
    \bottomrule
    \end{tabular}
    \tablefoot{Our statistical results are expressed in the form of the KL divergence and the relative deviation between the two distributions (FoM). These results include only L$_3$ level data with no outliers. The mass of the progenitor was estimated as an interval of equally probable values. We call NS the lowest end and BH the highest end.}
    \label{tab:results}
    \end{table}

    \subsection{Relative deviation from true value} 

    We introduced a second figure of merit (FoM) to quantify the accuracy of single-event parameter reconstruction. Given an estimated value, $S_i$, and the true value, $T_i$, of the parameter, $x$, we defined:
    \begin{equation} 
    \mbox{FoM}_x = \frac{1}{n}\sum_i^n \frac{|S_i - T_i|}{T_i}, 
    \end{equation} 
    where $n$ is the total number of events. This metric measures the average relative deviation between estimated and true values, expressed as a pure number. 
    
    The FoM complements the KL divergence. While KL divergence evaluates the overall accuracy of reconstructing the full distribution of values, the FoM focuses on individual event-level deviations. In this sense, the FoM serves as an indication of the relative error in estimating each parameter, and should be interpreted not as an absolute value but as an indication of the feasibility of parameter estimation given the available data. Results are shown in Table~\ref{tab:results}. Note that, similarly to the KL divergence, the FoM value exceeds 1 when the relative deviation is higher than 100$\%$.

\begin{figure*}
    \includegraphics[width=1\textwidth]{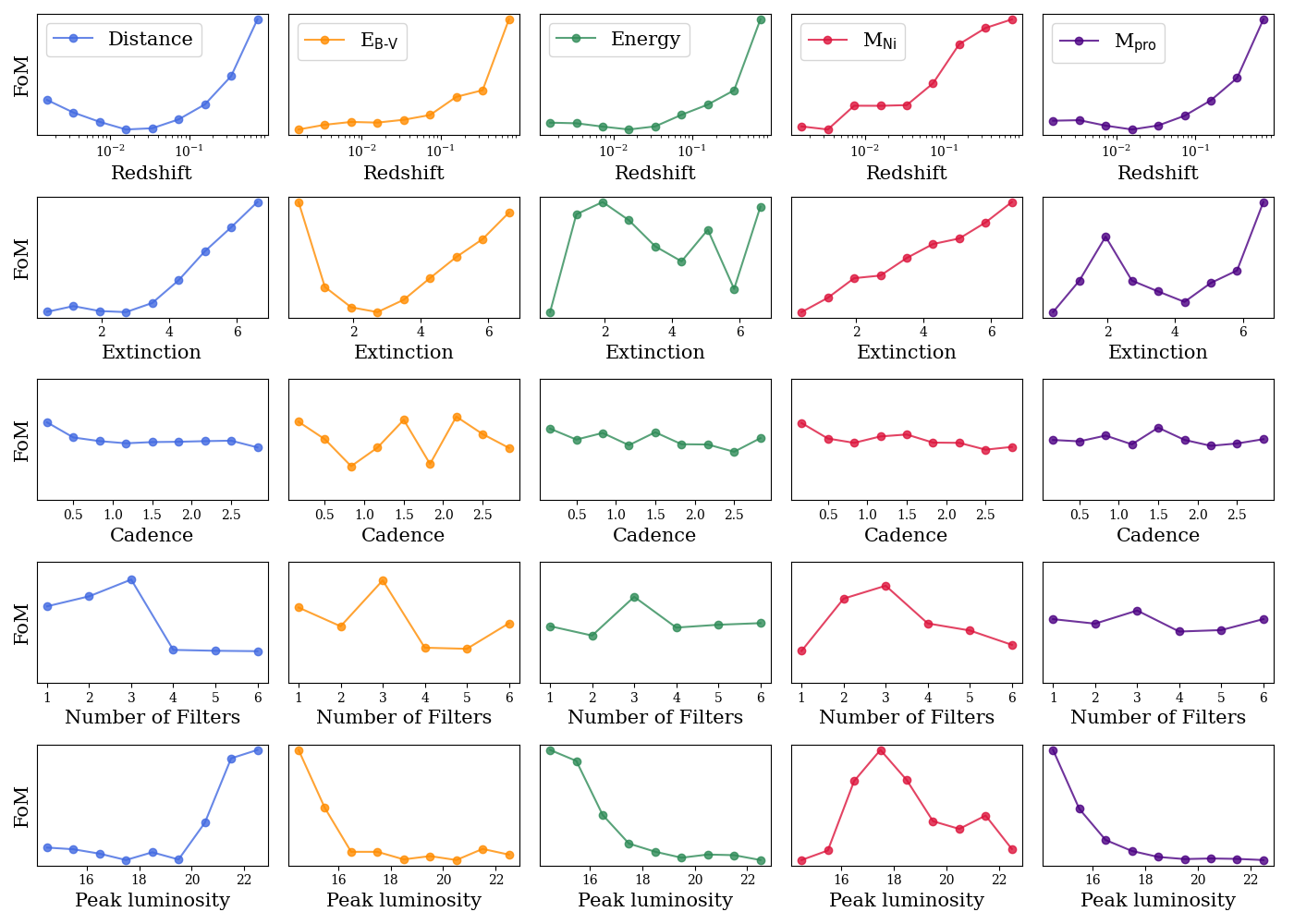}
    \caption{Relative deviation of distance (first column from the left), extinction (second column), energy (third column), mass of nickel (fourth column), and mass of progenitor (fifth column) from the true values in bins of redshift (first row from the top), extinction (second row), cadence (third row), number of filters (fourth row), and luminosity at peak (fifth row). Each point represents the relative deviation (FoM) for each bin, normalized by the number of events contained in each bin. The y axis is free-scaled to emphasize variations in deviation between bins rather than absolute values.
    }
    \label{fig:binning}
    \end{figure*}

    Binning the relative error in terms of redshift, extinction, cadence, number of filters, and peak luminosity can offer a valuable insight into how different observing strategies have an effect on the final reconstruction of the event. We created nine (six in the case of the number of filters) evenly distributed bins for each category (Fig.~\ref{fig:binning}). Each point represents the average FoM computed in every bin of each parameter, normalized by the number of events per bin, which allows for a direct comparison between bins that are populated by a very different number of events. The binning analysis allows one to get an idea of how the geometry of an event and the observing strategy may globally affect the parameter estimation. For instance, the FoM of all parameters increases with redshift and extinction and decreases with the number of filters. Notably, the observation cadence has little impact on parameter estimation. This is a direct consequence of using pre-interpolated datasets, which effectively eliminate differences between various cadence strategies. This highlights the importance of interpolation methods, such as the GP method, in handling different observing strategies during the LSST survey. By relying on such techniques, accurate parameter reconstruction remains possible even when an event is observed less frequently than others.

\section{Additional tests}

    In addition to the standard unsupervised analysis, we conducted two additional runs to address specific factors: disentangling the redshift-extinction degeneracy by fixing the extinction value and using unsaturated light curves. 
   
    \subsection{Breaking degeneracy}

    As was mentioned before, SN parameters exhibit a high degree of degeneracy, making it challenging to disentangle extinction and redshift without external information, such as spectral measurements. We simulated a scenario in which external data was available by fixing the extinction values, thereby reducing the degrees of freedom in the analysis.
    Our tests (Table~\ref{tab:results2}) indicate that fixing the extinction significantly improves parameter estimation, particularly for distance, as is evidenced by a notable reduction in KL divergence and lower deviation. Additionally, breaking this degeneracy benefits the estimation of explosion energy and progenitor mass, primarily due to the propagation of extinction errors. This issue largely affects SNe with missing data around the peak due to saturation. Furthermore, the number of outliers is substantially reduced, with only 3.7 percent of SNe excluded due to extreme energy and mass values. However, the estimation of nickel mass appears to be slightly less accurate in this scenario and the number of mass of nickel outliers remain high. This finding underscores the influence of other factors, such as the assumed model and the intricate interdependencies among parameters.

    \begin{table}
    \centering
    \caption{
    KL divergence and mean deviation (FoM) for the two additional runs.
    }
    \label{tab:peak}
    \begin{tabular}{l|cc|cc} 
    \toprule 
    Parameter               & D$_{KL}$\tablefootmark{a} & FoM\tablefootmark{a} & D$_{KL}$\tablefootmark{b} & FoM\tablefootmark{b}  \\  
    \midrule 
    Distance                  & 0.78    & 1.55  & 1.28 & 4.26   \\
    Extinction                & -       & -     & 0.27 & 1.27   \\
    Energy                    & 1.30    & 2.69  & 1.43 & 2.95    \\
    Mass of nickel            & 2.45    & 15.58 & 2.69 & 11.70  \\ 
    Mass of progenitor (NS)   & 0.64    & 0.77  & 0.56 & 0.69    \\ 
    Mass of progenitor (BH)   & 0.16    & 0.49  & 0.11 & 0.42    \\ 
    \bottomrule
    \end{tabular}
    \tablefoot{
    \tablefoottext{a}{Fixed extinction}; \tablefoottext{b}{unsaturated light curves.} }
    \label{tab:results2}
    \end{table}

    \subsection{Ignoring saturation} 

    We conducted an additional test by saving light curves from L$_2$ to L$_3$ without applying the saturation limit. Our results (Table~\ref{tab:results2}) indicate that, overall, parameter estimation improves when the light curves are unsaturated. Furthermore, this analysis produces only three progenitor mass outliers, underscoring the challenges posed by gaps in light curves, particularly in accurately determining energy, extinction, and progenitor mass. However, in this case, errors at low redshift are significantly higher than in the standard scenario, as are the errors at high peak luminosities (low magnitudes). Notably, at high luminosities (where saturation typically occurs) errors remain elevated, suggesting that these cases are inherently difficult to handle, regardless of whether the saturation limit is ignored. 
    
    \section{Discussion}

    We discuss the results from the analysis of the FoM, which highlight some crucial aspects to be taken into consideration when analyzing data to estimate SN progenitor parameters from the LSST survey.

    \subsection{Distance} 
    
    The distribution of the observed distance appears to be slightly above the limit of good consistency given by the KL parameter (D$_{KL}$ = 1.2). As is seen in Fig.~\ref{fig:comparison}, the two distributions are quite similar at medium distances, with some discrepancy particularly at $\sim$ 50 to 300 Mpc, with the highest distances ($\sim 500$ Mpc and more) not being reconstructed. This behavior is strongly exhibited when we bin the distance values in terms of redshift (first panel at the top left of Fig.~\ref{fig:binning}): the minimum of the FoM is found between redshift 0.01 and redshift 0.1. Going to lower redshifts, the distribution appears to increase almost linearly, while at higher redshifts it appears to increase almost exponentially. The same behavior can be seen in the extinction and peak luminosity bins as, typically, fainter objects with high extinction are the furthest. We showed how the mean error and the general distribution of distance can get significantly better when breaking the degeneracy with the extinction value. The current state of the art does not allow for good coverage of high-redshift SNe, limiting our reconstruction capability (as is shown in Fig.~\ref{fig:training}). Therefore, future observations at high distances might increase the available databases, allowing for a greater coverage, partially reducing the error at high redshift. 

    \subsection{Extinction}

    Along with redshift and distance, the extinction of a SN is a fundamental parameter on which all the other parameters depend. \texttt{CASTOR} estimates extinction, directly determining the $E_{B-V}$ parameter from the peak magnitude of a blue and visible filter. It then applies Cardelli's law following the prescriptions of \citet{McCall2004} to estimate absorption. However, if fewer than two filters are available without a gap around the maximum luminosity, we adopted a reference value assuming $R_V=3.1$, estimated using the \texttt{extinction} Python package \citep{Barbary2021} based on the available data. A clear distinction between these two methods is visible in Fig.~\ref{fig:comparison}: Cardelli's law, due to its reliance on a limited number of filters, produces a peak around $E_{B-V} = 0 - 1$, whereas the direct estimate, benefiting from the ability to use a greater number of filters more accurately, results in a peak around $E_{B-V} = 2 - 3$. However, neither method accurately reproduces high extinctions in distant SNe, exhibiting a similar behavior to that seen in other parameters when increasing redshift.

    The additional tests presented in Table~\ref{tab:results2} demonstrate that extinction estimation suffers from poor accuracy, particularly in cases of saturated SNe. Moreover, fixing the extinction leads to improved results for nearly all parameters, with a significant enhancement in distance estimation. Alternatively, when considering the unsaturated light curves, the relative error diminishes significantly, yielding a more consistent distribution. In both cases, these results emphasize the importance of external information (e.g., spectral coverage of the event) for accurately estimating extinction, as it can be the key factor determining whether other parameters are well or poorly reconstructed.

    \subsection{Energy} 

    Figure~\ref{fig:comparison} highlights a notable discrepancy between the observed and true energy distributions. While the true distribution is evenly spread in the range of $0.5-5 \times 10^{51}$ erg, the estimated energy distribution has a median value of $1 \times 10^{51}$ erg, with a tail of a few events reaching higher and lower values. This discrepancy is primarily driven by distance errors, particularly at extreme distances, as is shown in the first panel of Fig.~\ref{fig:binning}. These errors lead to an underestimation of the bolometric luminosity. Conversely, incorrect extinction estimates tend to overestimate luminosity, with the relative deviation increasing as extinction increases. The influence of these errors extends to other parameters as well. The accuracy of energy estimation improves with the number of available filters, suggesting that broader wavelength coverage plays a key role in reducing uncertainties. Additionally, as the peak luminosity increases, the relative deviation of the energy error decreases, further reinforcing the importance of robust observational constraints.
    
    In total, 14 percent of events are flagged as energy outliers and excluded from the observed distribution used to compute figures of merit. Among them, 86 percent are flagged as “saturated.” These findings underscore the significant impact of light curve saturation on accurately characterizing an event’s explosion energy. Lastly, when extinction is fixed or the saturation limit ignored, the estimation is more accurate and the number of outliers decreases significantly. 

    \subsection{Mass of nickel}

    The mass of nickel is the most critical parameter, due to instrumental limitations, dependencies, and the vast range of magnitudes involved. As this parameter is estimated from the linear decay of the bolometric curve, it is particularly sensitive to the magnitude limit of the instrument and the sampling in that region, in the same way as the energy is extremely sensitive to the saturation limit. Therefore, the mass of nickel is underestimated for SNe with at least one filter above the magnitude limit. Furthermore, nearly 17 percent of SNe in the L$3$ sample do not exhibit a linear decay phase at all, making it impossible to estimate their nickel mass and reducing the available statistical sample. The observed median nickel mass is $M_{Ni} = 10^{-3} M_\odot$, and when removing outliers beyond $\pm$ two orders of magnitude a strikingly high number of outliers remain: 31.8 percent in total, including 18.9 percent with at least one light curve under the magnitude limit. When accounting for both the outliers and the SNe without a measurable linear decay, the effective sample size is significantly reduced. Consequently, the final analysis was conducted using only two thirds of the original dataset.   
    Fig.\ref{fig:binning} clearly shows how the error in this parameter is highly dependent on the estimate of the distance, more than any other parameter, following more or less the same trend in all categories apart from the peak of luminosity. 

\begin{figure}
    \includegraphics[width=1\columnwidth]{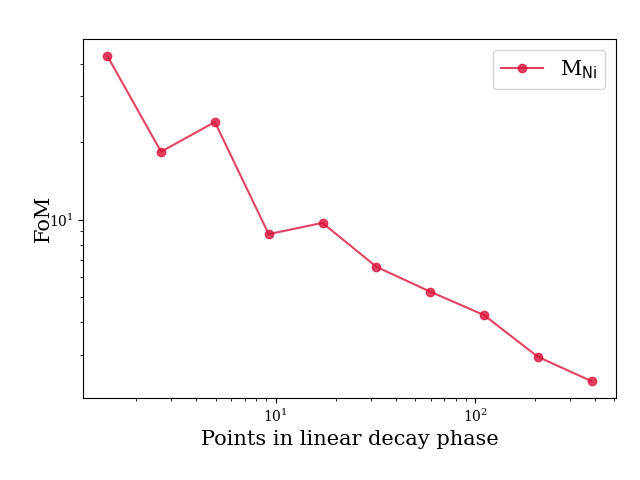}
    \caption{Mean deviation (FoM) of the observed mass of nickel from its true value binned in terms of the number of points available in the linear decay phase.}
    \label{fig:mni_points}
    \end{figure}

    We performed an additional test by binning the deviation of errors (FoM) in terms of the number of points available in the linear decay phase. This could help with identifying how dense the observations have to be performed in this region to allow for a good reconstruction of the parameter. We expect that this region carries the most weight in estimating the mass of nickel, as for the energy and the other parameters the region around the peak does. The final results are shown in Fig.~\ref{fig:mni_points}. The mean deviation decreases drastically from  FoM $\sim$ 20 to FoM $\sim$ 1, exhibiting how a good sampling of the linear decay can actually prevent one from badly evaluating the mass of nickel. On the other hand, in a realistic scenario, having more than tens of points sampled in the late stages of evolution is really unlikely, despite this being the best strategy to accurately characterize this parameter. 

    \subsection{Mass of the progenitor}

    The progenitor mass is a key parameter in modeling stellar evolution and understanding the mechanisms behind the explosion. It is one of the few parameters that can be constrained by independent observations and models, such as direct pre-explosion observations and hydrodynamical modeling. \texttt{CASTOR} estimates this parameter under the assumption of perfect mass conservation at the time of the explosion. As a result, the final value is presented as an interval of equally probable masses, ranging from the lowest to the highest possible remnant mass for this type of event. 
    
    The uncertainties in this parameter arise from three main sources: (1) propagation of errors in the energy estimate, leading to potential under- or overestimation of the ejecta mass and remnant mass; (2) inaccuracies in velocity estimation due to limitations in synthetic spectra modeling; and (3) model-dependent uncertainties. Overall, this parameter is the one most accurately reconstructed in this study. We excluded 8.5 percent of the total because they are outliers, of which 89 percent are events with at least one saturated light curve. For the remaining distributions, the KL divergence ranges from 0.74 to 0.18, corresponding to relative deviations of 0.84 and 0.52, respectively. Figure~\ref{fig:binning} illustrates this behavior. The trends in redshift and peak luminosity follow the same pattern as energy: higher redshifts make all parameters more challenging to reconstruct accurately, while peak luminosity exhibits an inverse relationship due to the systematic overestimation of energy at high luminosities and the different number of events per bin.

     \section{Conclusions}

     We analyzed LSST simulated light curves with \texttt{CASTOR v2.0} \citep{Simongini2024, castor_zenodo_2024}. The simulations, originally produced by \citet{Moriya2023}, are based on explosions of red supergiant stars modeled with the \texttt{KEPLER} code \citep{weaver1978presupernova, sukhbold2016core} and evolved using the radiative transfer code \texttt{STELLA} \citep{Blinnikov1998, Blinnikov2004, Baklanov2005, Blinnikov2006}. These synthetic events were then distributed in a three-dimensional space of redshift and extinction, resulting in a total of 22663 different SNe. Subsequently, we filtered the simulations on the basis of the limit magnitude, saturation limit, and number of remaining points, and interpolated with GP techniques the remaining light curves. The final dataset to analyze counts 6730 light curves. 
     \texttt{CASTOR} was used for every step of the analysis: (i) comparison with the light curves of a dataset of 106 SNe from the literature to identify a similar SN (reference SN); (ii) building synthetic spectra using the simulated light curves and the observed spectra from the reference SN; (iii) parameter estimation using GP interpolated light curves and synthetic spectra.   

     We estimated the distance, extinction, energy, mass of nickel, and progenitor mass, and compared the observed and the true distributions via two figures of merits: the KL divergence to evaluate how globally each distribution behaves compared to the true one and the relative mean deviation (FoM) to evaluate the accuracy of each observation, which we further binned in terms of redshift, extinction, cadence, number of filters, and peak luminosity (and the number of points in the linear decay for only the mass of nickel). Additionally, we performed two more tests to evaluate the relative weight of extinction and saturation.
     
     We highlight our major findings for each parameter: 

    \begin{enumerate}[i]

    \item Distance estimates are significantly influenced by two factors: the low reliability of the relationship between redshift and distance at small distances and technological limitations at large distances. However, in the intermediate range of $z = 10^{-2} - 10^{-1} $, distance reconstruction is more accurate, with a mean deviation of approximately 70$\%$. Overall, the distance parameter remains among the best reconstructed ones in our sample. In this work, we propose an alternative approach to distance estimation compared to the methods suggested in \cite{LSST2009}. That study recommends using either the expanding atmosphere method \citep{Schmidt1994} or the standardized candle method \citep{Hamuy2002}. While both techniques have been widely employed, they rely on external observational data. In contrast, \texttt{CASTOR} provides a way to estimate distances solely from light curves, eliminating the need for external information. This approach is similar to the photometric color method \citep{de2015hubble}, which constructs a Hubble diagram using corrected optical magnitudes derived solely from light curve data.
    
    \item Extinction estimates are strongly dependent on the saturation of light curves, as the accuracy decreases when less unsaturated filters are available. A good estimation of extinction may provide additional accuracy in the estimate of all the other parameters: this may be performed by external studies or with spectral information \citep[i.e., from the relative width of Na ID line;][]{Poznanski2012}.     

    \item The energy parameter relies heavily on bolometric luminosity, requiring at least four to six filters for accuracy. As redshift (and extinction) increase, energy estimates become less reliable, inversely to peak luminosity, while observation cadence has little impact. Sensitivity to the peak region makes energy estimates particularly affected by saturation limits: unsaturated cases yield better values, and fixing extinction helps reduce outliers.

    \item The nickel mass is a challenging parameter to estimate accurately, as it strongly depends on distance, extinction, luminosity, and the number of data points sampled at late times. We show that achieving a reliable nickel mass estimation requires observations in a sufficient number of filters or sampling the linear decay phase with up to $10^2$ data points.
    
    \item The mass of progenitor is a well-reconstructed parameter thanks primarily to the model presented in \texttt{CASTOR}, although it comes with high uncertainties regarding the mass of the remnant that are not considered in this work. Being dependent on the energy via the virial theorem, a better estimation of bolometric luminosity may provide a better reconstruction: this could be achieved with a higher number of filters. 

    \end{enumerate}

    This work explores the anticipated impact of the Vera C. Rubin Telescope, focusing on parameter estimation for core-collapse SNe. We examine the main challenges in this estimation process and discuss potential strategies to mitigate them. In particular, we demonstrate how \texttt{CASTOR} can effectively determine parameters without external data, achieving the highest accuracy at intermediate distances, in regions with relatively low absorption, and for SNe with medium peak luminosities. The optimal number of filters for reliable reconstruction is between four and six, while no specific constraints are required for observation cadence, thanks to the use of interpolation techniques.

    This work is particularly timely, as the LSST survey is expected to begin by the end of 2025. We anticipate that the current technological limitations will be overcome as soon as the first data become available, enabling a more precise characterization of CCSNe parameters. This will not only refine our understanding of these explosive events but also push the boundaries of transient astronomy, unveiling new phenomena and expanding our view of the dynamic Universe.

\bibliographystyle{aa}
\bibliography{bibliography}

\end{document}